Research paper

# Overcoming voltage fluctuation in electric vehicles by considering Al electrolytic capacitor-based voltage stabilizer

Ida Hamidah [a,*], Doni Fajar Ramadhan [a], Ramdhani Ramdhani [a], Budi Mulyanti [a], Roer Eka Pawinanto [a], Lilik Hasanah [b], Asep Bayu Dani Nandiyanto [b], Jumril Yunas [c], Andrivo Rusydi [d]

[a] *Faculty of Technology and Vocational Education, Universitas Pendidikan Indonesia, 40154 Bandung, Indonesia*
[b] *Faculty of Mathematics and Science Education, Universitas Pendidikan Indonesia, 40154 Bandung, Indonesia*
[c] *Institute of Microengineering and Microelectronics, Universiti Kebangsaan Malaysia, 43600 Selangor, Malaysia*
[d] *Department of Physics, National University of Singapore, Queenstown, Singapore*

**abstract**

Battery pack used in electrical vehicles (EVs) have developed fastly nowadays and are important due to its capability to stores electric charge. However, it is still found that there is an electrical voltage fluctuations due to sudden load changes or faults in the power system. To overcome this problem, the conventional strategy focuses on using voltage stabilizer with inductor-capacitor resonant circuits, leaving the problem of using capacitors at cryogenic temperatures. Herewith, we design a vehicle automatic voltage stabilizer (VAVS) operating at room temperature. The VAVS consists of nine main components, including resistors, capacitors, transistors, and diodes arranged with a specific composition to stabilize the voltage. By considering Al electrolytic capacitor, type of diode, resistor, and transistor, the VAVS device stabilizes the input voltage from 10-15V to obtain an output voltage of 12V. Then, by varying the engine rotation speed of vehicle, VAVS applied in vehicle battery supplies a constant voltage of about 12.1 V. The lowest line regulation analysis reveals a regulated voltage fluctuation of 0.17%/V. Our result shows that VAVS design works at room temperature, overcomes the voltage fluctuation and complies with the standard regulation of voltage stabilizers.

## 1. Introduction

In the development of motorized vehicle technology, the reliability of electrical energy is essential. As a midterm technology from fossil fuel vehicles to electric vehicles, hybrid electric vehicles have received wide research attention to make their efficiency. However, its efficiency is because most of today's motorized vehicles use the Electrical Control Unit (ECU) system, where there are a lot of electronic systems installed, thus requiring a stable supply of electrical energy (Çağatay Bayindir et al., 2011). The ECU is an electronic control device that functions as a regulator of actuator performance in vehicles that previously were only regulated by mechanical systems (Kleylein-Feuerstein et al., 2015). There are several types of ECU installed on the vehicle, including Engine Control Module (Dias et al., 2018), Powertrain Control Module (Di Ilio et al., 2021), Transmission Control Module (Wang et al., 2020), Brake Control Module (Dias et al., 2018; Zhu et al., 2021), Central Control Module (CCM), and Suspension Control Module (Theunissen et al., 2021). Some studies of ECU have been done, including implicit simulation technique full transient (Balakrishnan et al., 2017) and thermal management materials, which may apply to an automotive ECU (Mallik et al., 2011; Otiaba et al., 2011). A good supply of electrical energy is needed to support the performance of the ECU system in vehicles (Alam et al., 2019).

Nowadays, electrical vehicles (EVs) have developed fastly. One main component of an EV is the battery pack, which functions as electrical energy. It is still found that there is heat generation during the battery pack operation. Battery thermal management systems (BTMS) are needed to reduce the battery's temperature (Hamed et al., 2022; Karimi et al., 2022). Integrating solid-state thermoelectric refrigeration and heat pump systems with battery packs (Lyu et al., 2021), as well as using an evaporative cooling system with transcritical $CO_2$ cycles (Yin et al., 2022) can reduce the battery's temperature. In other side, due to their high energy density, lithium-ion batteries are frequently employed in electric vehicles (EVs). Until now, many studies have been done to improve the performance of lithium ion batteries. Some of them are to predict the whole-life-cycle state of charge (SOC) of lithium ion batteries using feedforward-long short-term memory (FF-LSTM) modeling method (Wang et al., 2022), multi-hidden layer long short-term memory (MHLSTM) neural network and suboptimal fading extended Kalman filtering (SFEKF) (Xie et al., 2023), and variable forgetting factor limited memory recursive least squares (VFF-LMRLS) method (Takyi-Aninakwa et al., 2023).

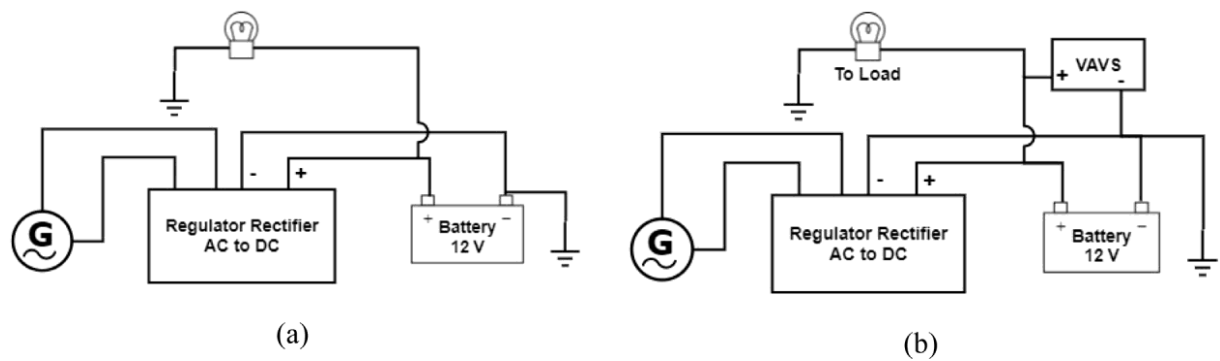

**Fig. 1.** (a) Regulated battery voltage system without VAVS, (b) Regulated battery voltage system with VAVS.

The supply of electrical energy to vehicles still causes problems, primarily due to fluctuations in electrical voltage as a consequence of sudden changes in load or faults in the power system (Hamidah et al., 2019). The voltage fluctuations can reduce service life and even cause electrical equipment operation's temporary or permanent failure, resulting in substantial economic losses (Su et al., 2019). Therefore, a voltage stabilizer is necessary to maintain a stable supply of voltage to the load and overcome fluctuations so that electrical equipment can be protected from over or under voltage. Some researchers have conducted in-depth studies on the characteristics of voltage stabilizers and their applications in various electrical and electronic systems. Some examples of voltage stabilizer research such as research on optimizing inductor-capacitor circuits (Kawashima et al., 2009), application of voltage stabilizers on modern materials (Englund et al., 2009), voltage doubler simulation using MATLAB (Balamurugan et al., 2014), ultra-capacitor-based voltage stabilizer (Villegas et al., 2015), automatic voltage regulator with an angle-based switching strategy (Batmani and Golpîra, 2019), capacitors design for voltage stabilizer (Farsadi et al., 2016), and Pareto analysis (Torreglosa et al., 2016) have contributed to improving the performance quality of the voltage stabilizer.

Even though all studies have been done to improve the performance quality of the voltage stabilizer, very few studies are conducted to apply voltage stabilizers to motorized vehicles. Thus, the design of the mobile voltage stabilizer device in this study is expected to be an alternative solution in maintaining electrical equipment from voltage fluctuations.

Fig. 1(a) shows the conventional electric vehicle power system installed without a VAVS. When the ignition is ON, current flows from the battery to the alternator. An electric current will pass through the stator coils, creating a magnetic field inside the alternator. When the engine rotates, the crankshaft and armature on the alternator rotate, and the electrical current is produced. The alternator's output is generally alternating current (AC), and a regulator/rectifier located inside the alternator is used to make the current become direct current (DC). When the engine rotates at high rotation, the electric current is also of a higher voltage—the charging system circuit is equipped with a regulator/rectifier to avoid an overvoltage. The output from the VAVS regulator will be used as energy to supply all electrical loads (Fig. 1(b)). Here, adding VAVS after the battery can stabilize the voltage from overvoltage charging and provide the electrical power that further prevents damage to other electrical components. The VAVS circuit is arranged in series with the battery so that it also turns off or does not function when the vehicle ignition is in the off position. If the VAVS are connected in parallel, the battery power will be sucked in by the VAVS when the vehicle is not in use.

In this paper, the performance of the VAVS in stabilizing the electrical vehicle system is discussed. As mentioned before, battery thermal management systems operate at cryogenic temperature; VAVS in this study are designed to operate at room temperature. This, of course, will provide more economic benefits. The VAVS parameters are designed and modeled based on the analysis of electronic circuits and characteristics of the capacitor, resistor, diode, and transistor using LTSpice software. Then, the performance of the VAVS modeling is validated by simulation and experiment. Modeling on VAVS is expected to guide engineers in designing an electrical system, facilitating their impact on the engineering design and its application in the industry (Schuk et al., 2022; Yildirim and Campean, 2020). The main contribution of this paper is that the research results provide a practical parametric design methodology for the VAVS, which is usually done empirically. In addition, LTSpice simulation done before making the VAVS can be an alternative solution to economically producing VAVS.

## 2. Experimental section

In this paper, the voltage stabilizer design aims to quantitatively improve the VAVS device's performance. The design and analysis are conducted in several steps, as shown in Fig. 2. This study begins with electrical circuit design. The performance of the designed system was then simulated using LTSpice by including parameters of a diode, capacitor, resistor, and transistor, and analyzed the output voltage carried out on the VAVS. The device was assembled in the motor vehicle, and VAVS performance tests were carried out to validate the analysis.

### 2.1. VAVS simulation using LTspice

The simulations were done to simplify the design of the circuit, minimize the risk of failure when the experiment is carried out, and minimize the estimated cost required to experiment. The software used to design and implement the simulation is LTspice XVII. LTspice XVII is an analog circuit simulation software tool.

Fig. 3 shows the schematic of the voltage stabilizer circuitry. When the electrical current from the battery flows to the vehicle's electrical system, it should pass through the capacitors $C_1$ and $C_2$ and then flows through resistor R. It serves to trigger the transistor Q, which functions as a sensor to tell whether the voltage of the VAVS is more or less. The excess of voltage will be stored by the $C_4$ and $C_5$ and the lack of voltage will be added by $C_3$, so the VAVS output voltage is expected to be constant at 12 V. Meanwhile, Meanwhile, a diode ($D_2$) and zener diode ($D_1$) in the VAVS circuit functions as a current rectifier. Different with diode which flown current in one direction, zener diode allows current to flow backwards when a certain set voltage is reached.

**Table 1**
The parameter of VAVS simulation using LTSpice.

| Description | Type/material | Details |
| --- | --- | --- |
| Polar capacitor | Al polymer; Al electrolytic | 33–150 µF |
| Nonpolar capacitor | Al polymer; Al electrolytic | 20–150 nF |
| Resistor | | Power rate: 0.25–2 W; Resistance: 1–1000 ▲ |
| Transistor | TIP3055; | |
| Zenner diode | Fast recovery | $V_{breakdown}$: 4.7–600 V; $I_{ave}$: 0.1–30 |

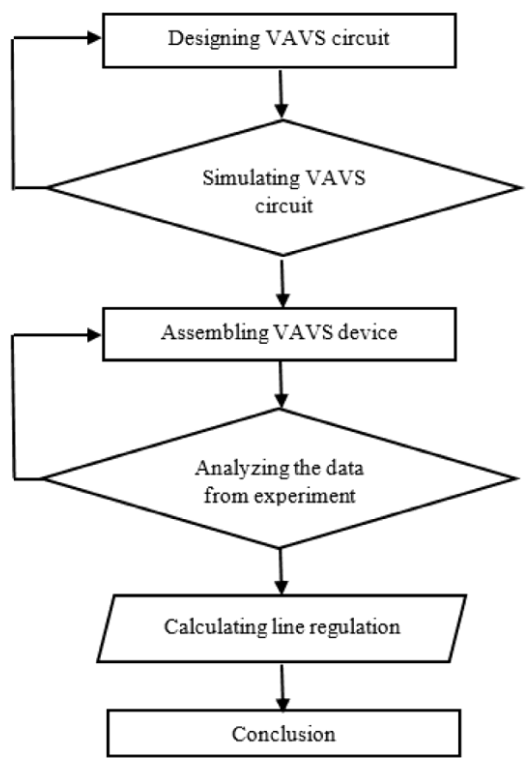

**Fig. 2.** VAVS design and analysis process flowchart.

LTSpice provides a large selection of parameters, and we select several commercially available parameters in the market. Each parameter is then varied to get the best results in voltage stabilization. Finally, a voltage stabilizer is made based on the best simulation results. Table 1 shows the parameter used in the VAVS simulation.

### 2.2. Circuit implementation

This stage takes input and output voltage measurement data from the VAVS assembled and implemented on the vehicle battery. The test is carried out by measuring the working voltage on the battery without using a VAVS device and a VAVS device. In measuring battery voltage without VAVS, measurements are made on the cable connecting the battery output and the vehicle's electrical system input. On the other hand, the battery voltage measurement regulated by the VAVS is carried out on the cable connecting the VAVS output and the vehicle's electrical system input, as shown in Fig. 4.

For testing, we first did not install VAVS on electric vehicles. We set the engine speed with varying values from 900–1500 rpm, and then the output voltage value was measured using a voltmeter. This output voltage value is then recorded as the input voltage if the electric vehicle is equipped with a VAVS. Next, we installed the VAVS on the electric vehicle with the diagram shown in Fig. 4. We did the same as in the test without VAVS by varying the engine speed and measuring the output voltage. The data from these measurements are shown in Fig. 6.

### 2.3. Line regulation analysis

One important characterization to determine the performance of an automatic voltage stabilizer is the Line Regulation which determines the ability of a device to maintain a constant output voltage despite changes to the input voltage. The line regulation is necessary when the input voltage source is unstable, resulting in significant variations in the output voltage (overvoltage or under voltage) (Aminzadeh, 2020). Line Regulation can be calculated based on the following equation:

$$\text{Line Regulation} = \frac{\Delta V_o}{\Delta V_i . V_o} . 100\% \tag{1}$$

where $\Delta V_i$ is the change in input voltage, $\Delta V_o$ is the change in output voltage, and $V_o$ is the output voltage.

The line regulation for an unregulated device is usually very high for most operations. However, this can be improved by using a voltage stabilizer. Low line regulation is always preferred (Peterchev and Sanders, 2006). According to IEEE 1159 Classification, a well-regulated device should have an overvoltage of at most 110 percent or 1.1 pu at the power frequency for a duration longer than 1 min Bhosale et al. (2018).

## 3. Result and discussion

### 3.1. Modeling and simulation using LTspice

Modeling and simulation using LTspice are done by changing the input voltage as a power source within the range between 10 to 15 V and by observing its signal property along the time. This modeling and simulation are carried out to determine how high the VAVS can produce the stabilized output. The results of the modeling and simulation are shown in Fig. 5.

The simulation begins by determining the VAVS electrical components with the parameters of each component available on the market. Then, the simulation is carried out for Vi of 10 V and varying the $V_{breakdown}$ of the Zenner diode while the other parameters are kept constant. The output voltage is constant by the time of 12.1 V for $V_{breakdown}$ of 600 V (Fig. 5a). The same simulation is done by changing the value of the resistor (Fig. 5b), and we get the $V_o$ constant value of 12.1 V for resister of 1 k▲. Then, the capacitance of a polar capacitor (Al electrolytic) was varied (Fig. 5c), and we get the Vo constant value of 12.1 V when 100 µf capacitance is used. Finally, we compare Al electrolytic and Al polymer capacitor. Fig. 5d shows that the Al electrolytic capacitor gives a constant output voltage of 12.1 V. Al electrolytic capacitors are frequently used in electronics applications if large energy storage capability is required (Ebel, 2023) and leakage current is not an important factor (Gudavalli and Dhakal, 2018).

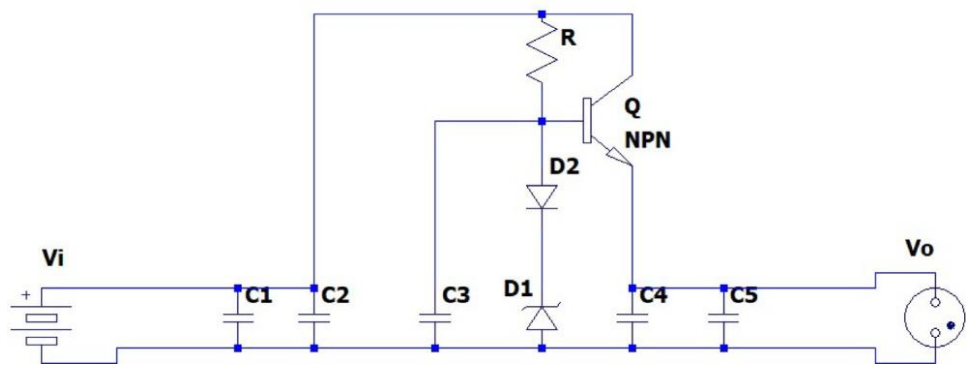

**Fig. 3.** VAVS design and analysis process flowchart.

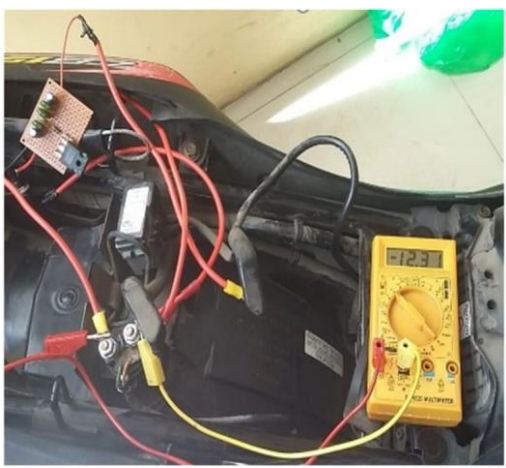
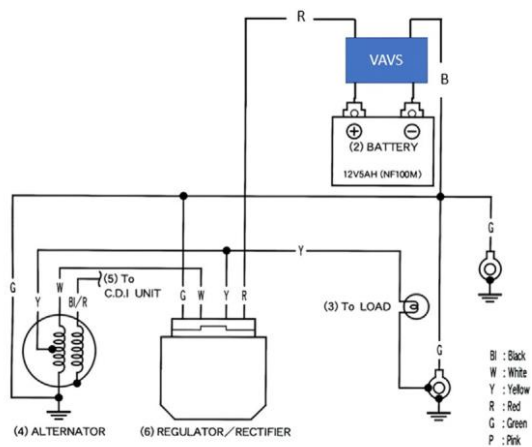

**Fig. 4.** (a) real experiment and (b) diagram of battery voltage measurement regulated by the VAVS.

**Table 2**
The parameter of VAVS simulation using LTspice.

| $V_i$ (V) | $V_o$ (V) |
|---|---|
| 10 | 12.1 |
| 11 | 12.1 |
| 12 | 12.1 |
| 13 | 11.98 |
| 14 | 12.2 |
| 15 | 12.15 |
| 10 | 12.1 |

Even though Al polymer capacitors exhibit considerably enhanced capacitance values and meager resistance (Yoo et al., 2015), Al electrolytic capacitors are responsible for the failure of power converters on a considerable scale (Niu et al., 2020). The output voltage against time in Fig. 5 shows that the VAVS can maintain a stable voltage without any disturbance.

Based on the results obtained in the first simulation, as shown in Fig. 5, the simulation is continued by changing the input voltage from 10 to 15 V. The results are summarized in Table 2. Table 2 shows that the VAVS can regulate the voltage with a stable 12 V. The highest output voltage is about 12.2 V, resulting in a line regulation of 102%, and this value is still below the maximum value according to the IEEE 1159 classification. Therefore, based on simulation and modeling using LTspice software, it can be concluded that the VAVS electronic circuit can regulate the voltage well.

### 3.2. VAVS implementation on motor vehicles

To test the performance of VAVS, then the experiment was carried out by comparing the output voltage of the vehicle battery regulated by VAVS and unregulated by VAVS. As we know, enhancing the engine rotation can improve the output voltage. The engine rotation in this experiment is set from 1500 (idling) to 9000 rpm.

The experimental results shown in Fig. 6 reveal that the battery regulated by VAVS can supply a stable average output voltage of about 12.1 V. This value is the standard for the vehicle battery voltage. On the other hand, an unregulated battery by VAVS produces an unstable output voltage when the engine rotation increases. The lowest line regulation of the regulated battery, voltage variation is about 0.17%/V when the engine rotation is approximately 2500 rpm. The average line regulation of regulated batteries is about 0.54%/V, which is still below the maximum value of the IEEE 1159 classification. These results show that the VAVS device works well and complies with standard regulations. The VAVS devices can work optimally because of the correct arrangement of the components, such as resistors, capacitors, diodes, transistors, and the proper composition. In addition, some researchers have also conducted experiments to stabilize the electric voltage in other areas, for example, determining the time delay of the contactless switching voltage stabilizing device in power supply system (Karimov and Bobojanov, 2020) and prototype voltage stabilizer using an optoelectronic contactless voltage relay, which can ensure reliable operation of the power supply network (Karimov et al., 2021). Finally, due to the stability of the voltage it provides, VAVS ensures that the overall performance of the system is good and offers reliable, robust, and cost-effective operation (Mokred et al., 2023; Leite Da Silva et al., 2000). Nevertheless, there are potential limitations to using VAVS in motorized vehicles because we did not test VAVS for long-term use. For further research, it is worth testing the lifetime of the VAVS.

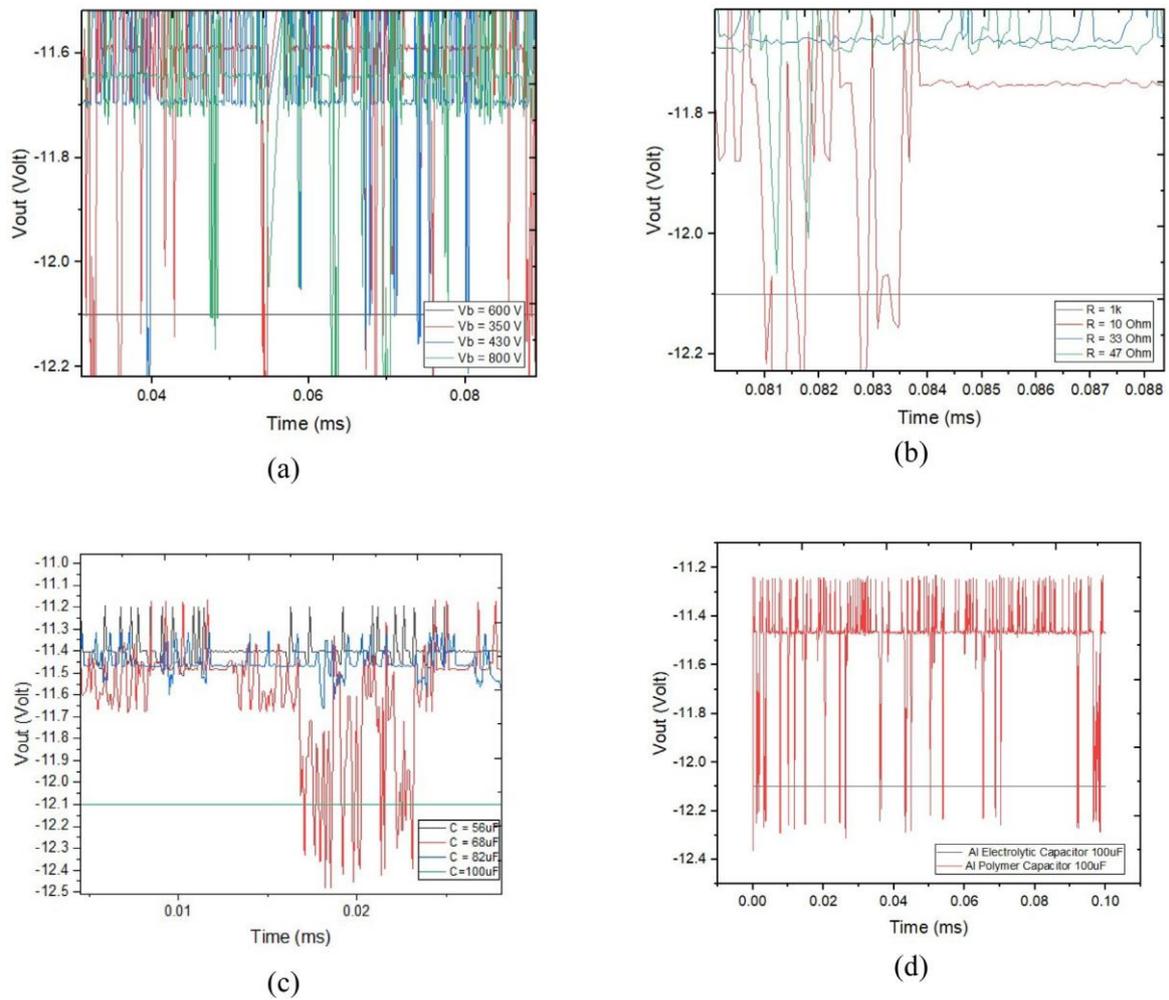

**Fig. 5.** The variation of output voltage against time is based on a variation of (a) diode, (b) resistor, (c) Al electrolytic capacitor, (d) Al electrolytic and Al polymer capacitor.

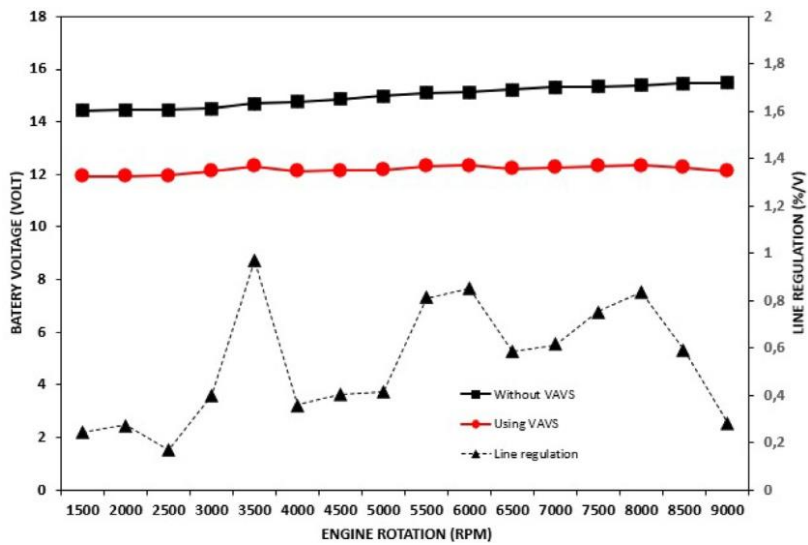

**Fig. 6.** Effect of engine rotations on the battery voltage output and line regulation characteristic on VAVS device.

## 4. Conclusion

The performance of VAVS has been studied using LTSpice simulation and implemented on a motor vehicle. Compared to conventional voltage stabilizer worked at cryogenic temperature, this VAVS more economics due to it works at room temperature. The characteristics of the VAVS, including the stabilization range and the overvoltage fluctuation of the battery, were discussed and compared with the measurement to protect the vehicle's electrical system from voltage fluctuation. The simulation showed that the VAVS device could stabilize the input voltage from 10–15 V to an output voltage of 12 V. The measurement of the battery output voltage showed that a constant voltage with an average voltage of about 12.1 V could be obtained. It can be concluded that through a correct arrangement of the components and compositions, the VAVS device can overcome the voltage fluctuations in the battery caused by changes in engine speed. Based on Line Regulation analysis, the lowest line regulation of the regulated battery variation is 0.17%/V with an average of 0.54%/V, showing that the VAVS device works very well and complies with the standard rule of voltage stabilizers.

**Declaration of competing interest**

The authors declare that they have no known competing financial interests or personal relationships that could have appeared to influence the work reported in this paper.

**Data availability**

The data that has been used is confidential

**Acknowledgments**

The authors would like to thank the Ministry of Education, Culture, Research, and Technology, the Republic of Indonesia, for funding this work under World Class Professor grant 2817/E4.1/KK.04.05/2021.

**Funding statement**

This study was supported by the Ministry of Education, Culture, Research, and Technology, the Republic of Indonesia through The World Class Professor grant number of 2817/E4.1/KK.04.05/2021 and Riset Kolaborasi Indonesia grant number of 0812/UN40/PT.01.02/2022.